\documentclass{article}

\usepackage[nonatbib,preprint]{neurips_2023_modified} 
\usepackage[numbers,sort&compress]{natbib}

\usepackage[utf8]{inputenc} 
\usepackage[T1]{fontenc}    
\usepackage{hyperref}       
\usepackage{url}            
\usepackage{booktabs}       
\usepackage{amsfonts}       
\usepackage{nicefrac}       
\usepackage{microtype}      

\usepackage{subcaption}%
\usepackage{graphicx}
\usepackage{colortbl}

\newcommand{\workshopName}{GAIED}

\newcommand{\textpapertitle}[1]{{\emph{#1}}}

\newcommand{\thrustO}{GAI\ensuremath{\rightarrow}ED}
\newcommand{\thrustC}{ED\ensuremath{\rightarrow}GAI}

\usepackage[usenames,dvipsnames]{xcolor}
\usepackage{hyperref}
\usepackage{enumitem}

\hypersetup{ %
	pdfauthor={GAIED},
	pdftitle={},
	pdfsubject={},
        pdflang={en},
	bookmarks,
	bookmarksnumbered,
	backref=page,
	pageanchor=true,
	plainpages=false,
	colorlinks=true,%
	linktocpage=true,
	citecolor=MidnightBlue,%
	filecolor=BrickRed,%
	linkcolor=NavyBlue,%
	urlcolor=NavyBlue 
}

\makeatletter

\newcommand{\Rmnum}[1]{\expandafter\@slowromancap\romannumeral #1@}
\makeatother

\title{Generative AI for Education (\workshopName):\\Advances, Opportunities, and Challenges\\
		\vspace{4mm}
		\normalsize{\textnormal{Overview of the \workshopName{} workshop at NeurIPS 2023 conference\thanks{\url{https://gaied.org/neurips2023/}}}}		
}

\author{
    Paul Denny\thanks{Authors are listed in alphabetical order.}\\
    University of Auckland\\
    \texttt{p.denny@auckland.ac.nz}
    \And
    Sumit Gulwani\footnotemark[2]\\
    Microsoft\\
    \texttt{sumitg@microsoft.com}
    \And
    Neil T. Heffernan\footnotemark[2]\\
    Worcester Polytechnic Institute\\
    \texttt{nth@wpi.edu}
    \And
    Tanja K{\"a}ser\footnotemark[2]\\
    EPFL\\
    \texttt{tanja.kaeser@epfl.ch}
    \And
    Steven Moore\footnotemark[2]\\
    Carnegie Mellon University\\
    \texttt{stevenjamesmoore@gmail.com}
    \And
    Anna N. Rafferty\footnotemark[2]\\
    Carleton College\\
    \texttt{arafferty@carleton.edu}
    \And
    Adish Singla\footnotemark[2]\\
    Max Planck Institute for Software Systems\\
    \texttt{adishs@mpi-sws.org}
}

\begin{document}

\maketitle

\begin{abstract}
This survey article has grown out of the \workshopName{} (pronounced ``guide'') workshop organized by the authors at the NeurIPS 2023 conference. We organized the \workshopName{} workshop as part of a community-building effort to bring together researchers, educators, and practitioners to explore the potential of \textbf{g}enerative \textbf{AI} for enhancing \textbf{ed}ucation. This article aims to provide an overview of the workshop activities and highlight several future research directions in the area of \workshopName{}.
\end{abstract}

%

\section{Introduction}
\label{sec.introduction}

Recent advances in generative AI, in particular deep generative and large language models like ChatGPT, are having transformational effects on the educational landscape. On the one hand, these advances provide unprecedented opportunities to enhance education by creating unique human-machine collaborative systems. For instance, these models could act as personalized digital tutors for students, as digital assistants for educators, and as digital peers to enable new collaborative learning scenarios. On the other hand, the advanced capabilities of these generative AI models have brought unexpected challenges for educators and policymakers worldwide. For instance, these advances have caused a chaotic disruption in universities and schools to design regulatory policies about the usage of these models and to educate both students and instructors about their strengths and limitations.

For us to fully realize these opportunities and tackle these challenges, it is time critical to build a community of researchers, educators, and practitioners that are ``multilingual'' with (a) technical expertise in the cutting-edge advances in generative AI, (b) first-hand experience of working with students in classrooms, and (c) know-how of building/deploying educational technology at scale. 
%
The goal of the \workshopName{} (pronounced ``guide'') workshop has been to foster such a multilingual community and bring together researchers, educators, and practitioners to explore the potential of \textbf{g}enerative \textbf{AI} for enhancing \textbf{ed}ucation. The workshop investigated these opportunities and challenges in education by focusing discussions along two thrusts:
\begin{enumerate} [label={\Roman*.}, parsep=2pt, leftmargin=*,labelindent=-6pt]
\item \textbf{\thrustO{}}: Exploring how recent advances in generative AI provide new opportunities to drastically improve state-of-the-art educational technology.
\item \looseness-1\textbf{\thrustC{}}: Identifying unique challenges in education caused by these recent advances and how to tackle them by bringing in desired safeguards along with technical innovations in generative AI.
\end{enumerate}


%

\section{Overview of the \workshopName{}@NeurIPS'23 Workshop Activities}
\label{sec.workshop_summary}

\looseness-1In this section, we provide an overview of the  \workshopName{} workshop organized at the NeurIPS'23 conference; full details are available on the \href{https://gaied.org/neurips2023/}{workshop website}. 
Given that this workshop bridged ideas in a central topic in machine learning (generative AI) and a key domain for societal impact (education), the workshop attracted considerable interest and engagement in the NeurIPS community.
%
%
The workshop was organized as a physical event, and we had up to $150$ attendees at the peak. The workshop program, including two poster sessions and a panel session, and the physical environment at the NeurIPS'23 venue provided ample opportunities for extensive discussions and networking among attendees.

\subsection{Topics of Interests}
As mentioned above, the workshop focussed on two thrusts,  each covering several topics of interest. These topics served as guidelines when selecting the speakers for invited talks and when accepting the contributed papers.

The topics in the \textbf{\thrustO{}} thrust focussed on leveraging recent advances in generative AI to improve state-of-the-art educational technology, including (i) exploring the capabilities of generative AI and large-language models in novel educational scenarios, e.g., personalized content generation and grading; (ii) exploring novel human-machine collaborative systems where generative models play different roles, e.g., as digital tutors, assistants, or peers; (iii) sharing viewpoints, novel ideas,  or field experiences about using generative AI in real-world educational settings. 
 
The topics in the \textbf{\thrustC{}} thrust focussed on tackling unique challenges in education by bringing in desired safeguards along with technical innovations in generative AI, including (i) developing novel prompting and fine-tuning techniques to safeguard the outputs of generative AI and large-language models against biases and incorrect information; (ii) developing novel safeguarding techniques to validate the authenticity of content, e.g., to determine whether an assignment was written by students or generated by models; (iii) sharing viewpoints, unique challenges, or field experiences about concerns among educators and policymakers in using generative AI.

\subsection{Accepted Papers and Poster Sessions}

\looseness-1In our call for papers and participation, we invited submissions of research papers reporting new results and position papers reporting new viewpoints or field experiences relevant to the workshop's topics of interest mentioned above.\footnote{Full details about the call are available at \url{https://gaied.org/neurips2023/index.html\#CFP}} For the reviewing process, we assembled a program committee of over $50$ reviewers, recruited both through invitations and through a special call for researchers with relevant experience to act as reviewer.\footnote{The final list of reviewers is available at 
\url{https://gaied.org/neurips2023/index.html\#Reviewers}
}
We had $48$ paper submissions in total. The reviewing process was double-blind, and each paper received at least three reviews from our program committee.

%

The final workshop program included $33$ accepted papers that were presented as part of two poster sessions. We also requested authors to prepare a $60$ second short lighting video to broaden the dissemination of their works. Accepted papers are publicly available as non-archival reports and are accessible on the workshop website along with the poster PDFs and video material submitted by authors.\footnote{Link to paper PDFs can be found at \url{https://gaied.org/neurips2023/index.html\#Papers}} 
In Section~\ref{sec.research_directions.current}, we provide an overview of the key research topics covered in these $33$ papers at the workshop.

\subsection{Invited Talks and Panel Session}

We invited a set of speakers with diverse backgrounds ranging from researchers with different expertise to industry practitioners and educators directly involved in educational activities. The invited speakers covered various topics of interest and achieved a balance across different perspectives and disciplines.  In total, the workshop had $6$ invited talks; each of these talks being about $20$ minutes long. 
The list of speakers, along with their talk titles, is provided below:
%

\begin{enumerate} [label={\arabic*.}, parsep=2pt, leftmargin=*,labelindent=2pt]
	\item \href{https://www.kristendicerbo.com/about-me}{Kristen DiCerbo} (Khan Academy):  \textpapertitle{Implementation of AI Tools in Education at Scale}. The talk discussed the development and research efforts behind Khanmigo, Khan Academy's GPT-4-powered tutor for students and assistant for teachers. In particular, the talk presented insights into Khanmigo, including the choice of models, approach to safety and security, and, most importantly, efforts to build learning science into the system so that it will help more learners learn more. References from the talk and related work by the speaker: \cite{Khanmigo}.
	%
	\item \href{https://glassmanlab.seas.harvard.edu/}{Elena Glassman} (Harvard University):  \textpapertitle{Generating and Revealing Structured Variation to Help Humans Learn Programming}. The talk discussed how AI-assisted programming tools are rapidly being integrated into CS classrooms and highlighted a key challenge of this rapid integration: reliance on AI assistance may hamper learning as learners may not be sufficiently cognitively engaged with the AI assistance. The talk also presented recent works on cultivating cognitive engagement through the use of aligned differences based on the design implications of various theories of human concept learning and related psychological studies. References from the talk and related work by the speaker:   \cite{DBLP:conf/iui/GajosM22,DBLP:conf/chi/Vaithilingam0G22,DBLP:conf/chi/Glassman0HK18,DBLP:conf/uist/YanKH0G22}.
	%
	\item \looseness-1 \href{http://www.hkeuning.nl/}{Hieke Keuning} (Utrecht University):  \textpapertitle{Learning Programming in the Era of Generative AI}. The talk discussed how the rise of generative AI and LLMs is changing the nature of programming and, consequently, the need to teach programming at different levels of abstraction with the need for different skills. The talk presented opportunities to use LLMs to support students' learning by generating formative feedback, explanations, and hints for programming tasks. The talk further highlighted challenges that LLM-based support may not be effective directly as its usefulness depends greatly on the learning goals and context. References from the talk and related work by the speaker: \cite{DBLP:conf/ace/Finnie-AnsleyDB22,DBLP:conf/iticse/Prather00BACKKK23,DBLP:journals/corr/abs-2310-05998,DBLP:journals/corr/abs-2209-14876,DBLP:journals/corr/abs-2311-05943,DBLP:conf/ace/RoestKJ24,DBLP:journals/corr/abs-2309-00029,denny2024computing}.
	%
	\item \href{https://tobiaskohn.ch/}{Tobias Kohn} (Karlsruhe Institute of Technology):  \textpapertitle{Hopes and Fears: Should We Use AI in Schools and Will It Revolutionise How We Think About Education?} The talk discussed how generative AI in education has been a highly polarising issue, with some welcoming the new opportunities while others warning of the risks of how students may utilize generative AI. The talk also provided perspectives on what it would mean to include the ``correct'' use of generative AI in our curricula. One of the key messages is that generative AI must not replace crucial human interaction in education. References from the talk and related work by the speaker:  \cite{DBLP:conf/iticse/Prather00BACKKK23,DBLP:conf/icer/PhungPCGKMSS22,edm23-pyfixv}
	%
	\item \href{https://stanford.edu/~cpiech/bio/index.html}{Chris Piech} (Stanford University):  \textpapertitle{Generative AI for Joyful Education}. The talk presented an overview of different AI and machine learning algorithms for enhancing education, covering a decade of research from the emergence of deep knowledge tracing in 2015 to recent developments in generative AI. One of the key messages is that we drastically undertilize human potential, as the magnitude of people who want to teach is roughly proportional to the magnitude of people who want to learn. The talk also highlighted new grand challenges and research directions in leveraging generative AI for education by developing AI models to augment teacher training, make learners generate new content, create curiosity among learners, and spark joy. References from the talk and related work by the speaker:  \cite{DBLP:conf/nips/PiechBHGSGS15,DBLP:conf/aaai/WuMGP19,DBLP:conf/nips/NieBP21,DBLP:conf/edm/WuDDPG20,DBLP:conf/lak/KimP23,DBLP:conf/lats/WangSLP17,DBLP:conf/sigcse/PiechMJS21,edm22-ai-teacher-test,DBLP:conf/lats/MarkelOLP23}
	\item \href{https://web.eecs.umich.edu/~wangluxy/}{Lu Wang} (University of Michigan):  \textpapertitle{Personalized Writing Learning Assistant using Natural Language Processing: Opportunities and Challenges}. The talk discussed ongoing efforts to leverage LLMs to build personalized argumentative writing assistants. The talk also presented recent works on building human-NLP collaborative tools to help educators create active learning opportunities at scale. The talk further highlighted challenges associated with employing LLMs for educational support because of hallucinations, outdated knowledge, the lack of complex reasoning abilities,  and the lack of creativity. References from the talk and related work by the speaker:  \cite{DBLP:conf/chi/LuFH0W23,DBLP:conf/naacl/WangFHW22,DBLP:conf/acl/Hua022,DBLP:conf/emnlp/ZhangKLLL023,DBLP:journals/corr/abs-2310-19208}.
\end{enumerate}

In addition to these invited talks, the speakers also participated in a panel session of $60$ minutes duration. In Section~\ref{sec.research_directions.future}, we provide an overview of the research themes discussed in the panel session. The video recordings of these invited talks and the panel session are publicly available.\footnote{Link to video recordings can be found at \url{https://gaied.org/neurips2023/index.html\#Speakers}}



\subsection{Diversity Statement}
We put a substantial effort into ensuring diversity at the workshop, including diversity among speakers and organizers in different aspects:
%
\begin{itemize}[parsep=2pt, leftmargin=*,labelindent=0pt]
    \item \textbf{Diversity in viewpoints, thinking, and expertise}: The list of speakers and organization team included people with diverse backgrounds and expertise who brought various viewpoints and thinking from varying experiences with generative AI for education. On the one hand, this included researchers in human-computer interaction, learning sciences, natural language processing, machine learning, and program synthesis. On the other hand, this included industry practitioners and educators directly involved in educational activities.
    \item \textbf{Diversity in demographics}: The list of speakers spanned $6$ institutes across $3$ countries and $4$ nationalities. The organization team spanned $7$ institutes across $4$ countries and $4$ nationalities.
    \item \textbf{Diversity in gender}: The list of speakers included $4$ women among $6$ speakers. The organization team included $2$ women among $7$ members.
    \item \textbf{Diversity in seniority}: The list of speakers included $4$ early career researchers from academia, $1$ senior researcher from academia, and $1$ senior member from industry. The organization team included $1$ Ph.D. student, $2$ early career researchers from academia, $3$ senior researchers from academia, and $1$ senior member from industry.
\end{itemize}

We continued to spend substantial outreach efforts throughout the workshop organization to encourage diverse participants to attend the workshop.
%

%

\section{Research Topics and Directions in \workshopName{}}
\label{sec.research_directions}
In this section, we provide an overview of popular research topics and highlight key research directions in the area of \workshopName{}.

\subsection{Popular Research Topics}
\label{sec.research_directions.current}

\begin{figure*}[h!]
    \centering
    \includegraphics[trim={0 1.2cm 0 0.6cm},clip,width=0.9\textwidth]{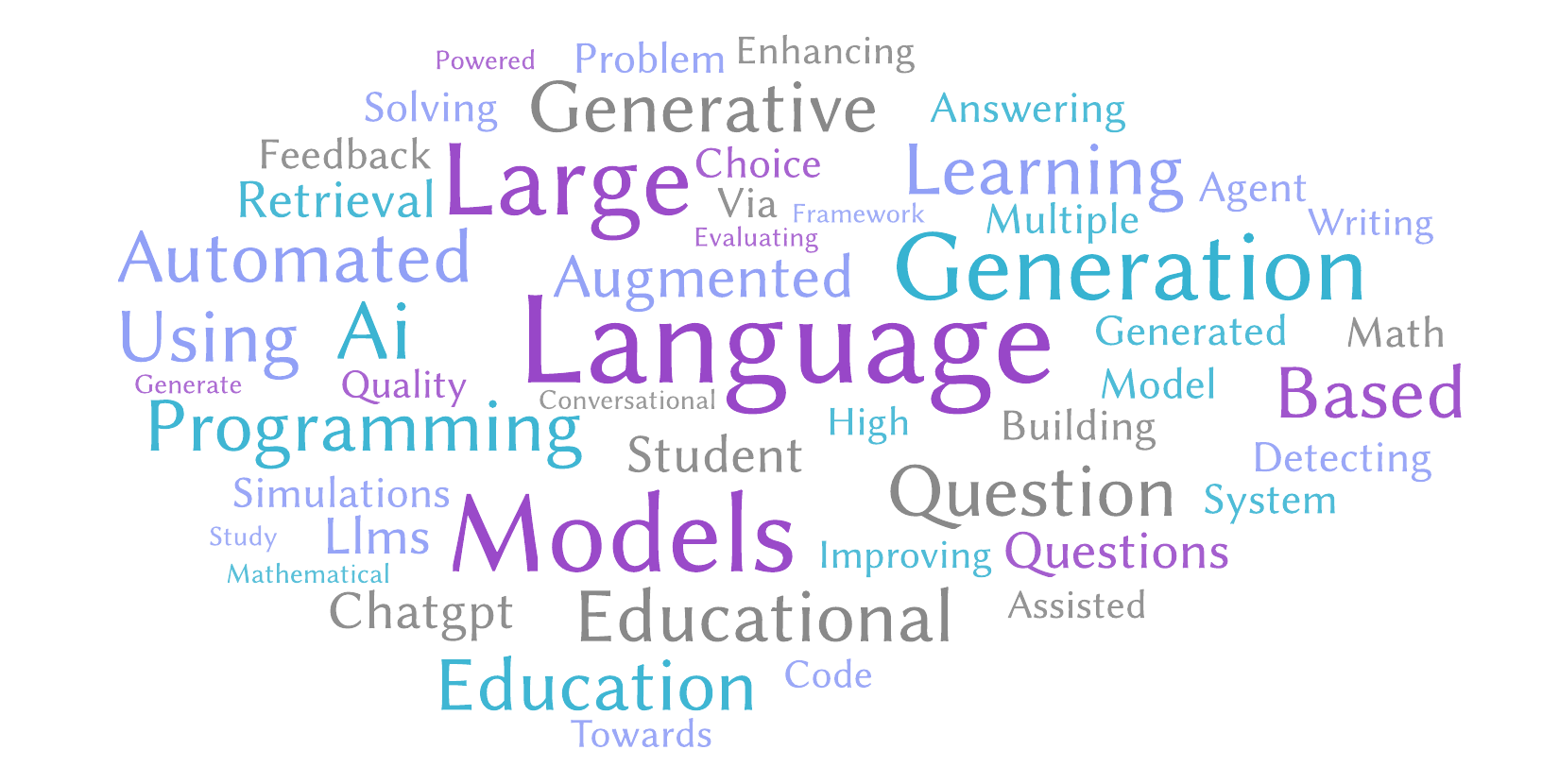}
    \caption{Word cloud created from titles of the $33$ accepted papers at the workshop.}
    \label{fig:word-cloud}
\end{figure*}

\looseness-1Here, we provide an overview of the popular research topics based on $33$ accepted papers at the workshop.\footnote{These papers capture a timely snapshot of popular topics in the community -- the workshop served as one of the first venues covering research conducted after the release of OpenAI's GPT-4 model and the availability of open-access models like Meta's LLama-2 model.} Figure~\ref{fig:word-cloud} illustrates a word cloud from titles of the papers, highlighting popular technologies, application domains, and concrete educational scenarios. Figure~\ref{fig:workshop-papers} summarizes the papers along different dimensions, including the application domain and educational scenario considered, models used, and specific learning/training techniques employed (e.g., in-context learning/prompting, pre-training, and fine-tuning).

\begin{figure*}[t!]
    \centering
    \scalebox{0.8}{
    \setlength\tabcolsep{1.5pt}
    \renewcommand{\arraystretch}{1.5}  
    \hspace{-14mm}
    \begin{tabular}{r || l | l | l }
        \toprule 
        \textbf{Paper}  & \textbf{Application Domain and Educational Scenario} & \textbf{Models} & \textbf{Techniques}\\
        \midrule
        $3$ \cite{neurips2023gaied_3_guo} & Detecting machine generated text to avoid misuse in academic settings & GPT-4, GPT-3.5 & In-context \\
        $4$ \cite{neurips2023gaied_4_jahanbakhsh} & Generating answers and explanations for different academic subjects & GPT-4, GPT-3.5 & In-context \\
        $5$ \cite{neurips2023gaied_5_koutcheme} & Generating program repair for Python programming education & StarCoder, CodeGen  & Fine-tuning \\
        %
        $7$ \cite{neurips2023gaied_7_padurean} & Generating new tasks for elementary-level visual programming & LSTM models, GPT-4  & New architecture \\
        %
        $8$ \cite{neurips2023gaied_8_lee} & Generative student agents for teacher training in elementary schools & GPT-4 & In-context \\
        $9$ \cite{neurips2023gaied_9_blobstein} & Generating Q/A across different cognitive levels for 8th-grade science & GPT-3.5 & In-context \\
        $10$ \cite{neurips2023gaied_10_mcnichols} & Generating distractors and feedback for math multi-choice questions & GPT-4, GPT-3.5 & In-context \\
        $11$ \cite{neurips2023gaied_11_zhang} & Conversational Q/A for educating patients about their clinical visits & GPT-4 & In-context \\
        $12$ \cite{neurips2023gaied_12_hayat} & LLMs for early forecasting of student performance in MATLAB course & FLAN-T5 & Fine-tuning \\
        $14$ \cite{neurips2023gaied_14_sonkar} & Generating synthetic student-teacher dialogues for physics & GPT-4, LLaMA-2 & In-context, fine-tuning \\
        $15$ \cite{neurips2023gaied_15_savelka} & Classifying student help requests in programming courses & GPT-4, GPT-3.5 & In-context, fine-tuning \\
        $16$ \cite{neurips2023gaied_16_pielage} & Generating images for virtual reality based clinical education & Diffuson models & New architecture \\
        $17$ \cite{neurips2023gaied_17_hwang} & Generating multi-choice questions aligned with Bloom's taxonomy & GPT-3.5, RoBERTa & In-context \\
        $18$ \cite{neurips2023gaied_18_fawzi} & Generating educational questions using small language models & T5-base, T5-small & Pre-training, fine-tuning \\
        $19$ \cite{neurips2023gaied_19_han} & New dataset of student-ChatGPT dialogues in EFL writing education & GPT-3.5 & In-context \\
        $21$ \cite{neurips2023gaied_21_anand} & Evaluating LLMs on high-school mathematical problem solving tasks & LLaMA-2, MAmmoTH & Fine-tuning \\
        $26$ \cite{neurips2023gaied_26_oli} & Investigating behavior of LLMs when generating code explanations & GPT-4, GPT-3.5, LLaMA-2 & In-context \\
        $27$ \cite{neurips2023gaied_27_lira} & Generating stories in Chilean Spanish to assist students in writing & GPT-3, BLOOM, GPT-2 & Fine-tuning \\
        $30$ \cite{neurips2023gaied_30_ghadban} & Retrieval-augmented generation using LLMs to enhance healthcare education & GPT-4 & In-context \\
        $31$ \cite{neurips2023gaied_31_asthana} & Generating comprehensive lecture metadata and questions from lecture videos & GPT-4 & In-context \\
        $32$ \cite{neurips2023gaied_32_zamfirescu-pereira} & Conversational programming and interactive support in CS courses & GPT-4 & In-context \\
        $33$ \cite{neurips2023gaied_33_bailis} & Generating mini-puzzles and images to aid language learning & PaLM, Imagen & In-context\\
        $34$ \cite{neurips2023gaied_34_choi} & Assisting teachers in Sierra Leone with LLM-powered chatbot via WhatsApp & GPT-3.5 & In-context \\
        $35$ \cite{neurips2023gaied_35_abdelghani} & Framework for introducing pedagogical transparency in GAI applications & n/a & n/a \\
        $38$ \cite{neurips2023gaied_38_schmucker} & Conversational tutoring systems with LLM agents via learning-by-teaching & GPT-4 & In-context \\
        $39$ \cite{neurips2023gaied_39_bulusu} & Producing graphs from natural language commands for mathematical pedagogy & GPT-4 with SymPy & In-context \\
        $40$ \cite{neurips2023gaied_40_levonian} & Interactive QA to support conceptual discussion of high-school math concepts & GPT-3.5 & In-context \\
        $41$ \cite{neurips2023gaied_41_lekan} & Personalized advising for undergraduate major recommendation in universities & GPT-4, GPT-3.5 & In-context \\
        $43$ \cite{neurips2023gaied_43_xu} & Retrieving exercises relevant to user queries for online language learning & GPT-4, mBERT & In-context, fine-tuning \\
        $44$ \cite{neurips2023gaied_44_bhandari} & IRT evaluation of LLMs' ability to generate math assessment questions & GPT-4 & In-context \\
        $46$ \cite{neurips2023gaied_46_sahai} & Generating feedback for high-school programming assignments & GPT-4, GPT-3.5 & In-context \\
        $47$ \cite{neurips2023gaied_47_roy} & Detecting educational content in videos for kindergarten literacy and maths & BLIP-2, Whisper & Supervised training\\
        $48$ \cite{neurips2023gaied_48_acun}  & Enabling conversations with historical scientists for history education & GPT-3.5 & In-context\\        
        \bottomrule
    \end{tabular}
    }
    \caption{Summary of the $33$ accepted papers at the workshop along different dimensions, including the application domain and educational scenario considered, models used, and specific learning/training techniques employed.}
    \label{fig:workshop-papers}    
\end{figure*}


\subsection{Key Research Directions}
\label{sec.research_directions.future}

Next, we highlight a few key research directions based on questions and discussions in the panel session at the workshop. 

\looseness-1\paragraph{Emerging curricula and skill sets.} Generative AI is bringing in transformational effects on the educational landscape, quickly changing how we think about education, from how we teach to what skills students need. For instance, generative models like GPT-4 can provide solutions to the questions typical in many undergraduate curricula. In light of these changes, it is time-critical for the community to jointly explore what this means for instructors and students, and how curricula should evolve to keep up with these advances. Some of the key questions along this theme are:
\begin{itemize}[parsep=2pt, leftmargin=*,labelindent=0pt]
    \item  \looseness-1How do teaching practices, instructors' roles, and curricula need to adapt given the ubiquitous availability of generative AI models and their capabilities to solve questions in undergraduate curricula? 
    \item How might the ubiquity of AI in generating content challenge current assessment methods, and what innovative strategies could be employed to fairly and effectively evaluate student learning? 
    \item Are there any risks that the usage of generative AI in educational settings could homogenize critical thinking or creativity among students, and what measures can be taken to prevent this?
    \item How can we utilize generative AI to personalize education at scale, and what curricular changes are necessary to support individualized learning journeys?
\end{itemize}

\paragraph{Diversity, equity, inclusion, and accessibility.} With the introduction of AI tools in education, it is crucial to ensure that these tools benefit everyone and increase equity in education. However, there is an increasing concern that the current trajectory of AI in education could lead to a wider gap between institutions/individuals that can afford these technologies and those that cannot. Consequently, it is crucial for us to explore AI's potential in either closing or widening the gap in educational opportunities, and discuss ways to ensure these technologies are accessible and fair to all learners. Some of the key questions along this theme are:
\begin{itemize}[parsep=2pt, leftmargin=*,labelindent=0pt]
    \item What measures can be taken to increase the accessibility of AI educational tools to underprivileged or underrepresented communities in order to bridge the digital and educational divide?
    \item How can the algorithms behind generative AI be audited for fairness and transparency to ensure that all students benefit equally from these technologies?
    \item What are the ethical considerations when using AI to track and analyze student performance data, particularly concerning the privacy and agency of students from vulnerable populations?
    \item How can we combat the risk where AI systems, mostly developed in technologically advanced countries, are imposed upon educational institutions in less developed regions?
\end{itemize}

\paragraph{Metrics and evaluation.} The challenge with new AI technologies in education doesn't stop with their development; it continues into how we deploy and assess their effectiveness in real classrooms. For instance, one of the challenges for AI and machine learning researchers developing new generative models for educational use is what automated metrics, objectives, and benchmarks to use that align the model's performance with educational goals. To ensure that developed generative educational technologies will achieve the desired educational goals, the community must connect the dots between basic AI research and classroom outcomes, creating ways to reliably measure the impact of AI on student learning over time. Some of the key questions along this theme are:
\begin{itemize}[parsep=2pt, leftmargin=*,labelindent=0pt]
     \item What specific benchmarks and standardized metrics should be developed to rigorously test and compare the effectiveness of various generative AI tools developed for educational applications?   
    \item How can we improve the pipeline from research idea, to technical development, to deployment and evaluation with real students?
    \item Given the constant development and changes in models like GPT-4 and Gemini that big corporations own, how should we ensure reproducibility and comparison of techniques developed over time?  
    \item How can we establish longitudinal studies to gauge the long-term effects of generative AI on students' critical thinking, creativity, and ability to learn independently?
\end{itemize}


\section{Conclusions}
\label{sec.conclusions}

Generative AI for education is an important, time-critical area for researchers, educators, and practitioners: ongoing advances in generative AI will continue providing unprecedented opportunities to enhance education, and unique challenges in education will require further technical innovations in generative AI. The talks, papers, and discussions at the \workshopName{}@NeurIPS'23 workshop highlighted excitement in the community around the main areas covered in this article, and the diverse array of perspectives as well as panelist comments demonstrated the importance of drawing on ideas from multiple disciplines, including (but not limited to) human-computer interaction, learning sciences, natural language processing, machine learning, and program synthesis. This need for multiple perspectives and the unique challenges raised by educational applications highlight the need for continued fostering of community in the area of \workshopName{}.

%

\bibliography{main}

\begin{thebibliography}{10}

\bibitem{Khanmigo}
Khan Academy.
\newblock Khanmigo.
\newblock \url{https://www.khanacademy.org/khan-labs}, 2023.

\bibitem{DBLP:conf/iui/GajosM22}
Krzysztof~Z. Gajos and Lena Mamykina.
\newblock {D}o {P}eople {E}ngage {C}ognitively with {AI}? {I}mpact of {AI}
  {A}ssistance on {I}ncidental {L}earning.
\newblock In {\em Proceedings of the International Conference on Intelligent
  User Interfaces (IUI)}, pages 794--806, 2022.

\bibitem{DBLP:conf/chi/Vaithilingam0G22}
Priyan Vaithilingam, Tianyi Zhang, and Elena~L. Glassman.
\newblock {E}xpectation vs. {E}xperience: {E}valuating the {U}sability of
  {C}ode {G}eneration {T}ools {P}owered by {L}arge {L}anguage {M}odels.
\newblock In {\em Proceedings of the CHI Conference on Human Factors in
  Computing Systems (CHI)}, 2022.

\bibitem{DBLP:conf/chi/Glassman0HK18}
Elena~L. Glassman, Tianyi Zhang, Bj{\"{o}}rn Hartmann, and Miryung Kim.
\newblock {V}isualizing {API} {U}sage {E}xamples at {S}cale.
\newblock In {\em Proceedings of the CHI Conference on Human Factors in
  Computing Systems (CHI)}, 2018.

\bibitem{DBLP:conf/uist/YanKH0G22}
Litao Yan, Miryung Kim, Bjoern Hartmann, Tianyi Zhang, and Elena~L. Glassman.
\newblock {C}oncept-{A}nnotated {E}xamples for {L}ibrary {C}omparison.
\newblock In {\em The Annual Symposium on User Interface Software and
  Technology (UIST)}, 2022.

\bibitem{DBLP:conf/ace/Finnie-AnsleyDB22}
James Finnie{-}Ansley, Paul Denny, Brett~A. Becker, Andrew Luxton{-}Reilly, and
  James Prather.
\newblock {T}he {R}obots {A}re {C}oming: {E}xploring the {I}mplications of
  {O}pen{AI} {C}odex on {I}ntroductory {P}rogramming.
\newblock In {\em Australasian Computing Education Conference (ACE)}, 2022.

\bibitem{DBLP:conf/iticse/Prather00BACKKK23}
James Prather, Paul Denny, Juho Leinonen, Brett~A. Becker, Ibrahim Albluwi,
  Michelle Craig, Hieke Keuning, Natalie Kiesler, Tobias Kohn, Andrew
  Luxton{-}Reilly, Stephen MacNeil, Andrew Petersen, Raymond Pettit, Brent~N.
  Reeves, and Jarom{\'{\i}}r Savelka.
\newblock {T}he {R}obots {A}re {H}ere: {N}avigating the {G}enerative {AI}
  {R}evolution in {C}omputing {E}ducation.
\newblock In {\em Proceedings of the Working Group Reports on Innovation and
  Technology in Computer Science Education (ITiCSE-WGR)}, 2023.

\bibitem{DBLP:journals/corr/abs-2310-05998}
Johan Jeuring, Roel Groot, and Hieke Keuning.
\newblock {W}hat {S}kills {D}o {Y}ou {N}eed {W}hen {D}eveloping {S}oftware
  {U}sing {C}hat{GPT}? ({D}iscussion {P}aper).
\newblock {\em CoRR}, abs/2310.05998, 2023.

\bibitem{DBLP:journals/corr/abs-2209-14876}
Jialu Zhang, Jos{\'{e}} Cambronero, Sumit Gulwani, Vu~Le, Ruzica Piskac,
  Gustavo Soares, and Gust Verbruggen.
\newblock {R}epairing {B}ugs in {P}ython {A}ssignments {U}sing {L}arge
  {L}anguage {M}odels.
\newblock {\em CoRR}, abs/2209.14876, 2022.

\bibitem{DBLP:journals/corr/abs-2311-05943}
Paul Denny, Juho Leinonen, James Prather, Andrew Luxton{-}Reilly, Thezyrie
  Amarouche, Brett~A. Becker, and Brent~N. Reeves.
\newblock {P}rompt {P}roblems: {A} {N}ew {P}rogramming {E}xercise for the
  {G}enerative {AI} {E}ra.
\newblock {\em CoRR}, abs/2311.05943, 2023.

\bibitem{DBLP:conf/ace/RoestKJ24}
Lianne Roest, Hieke Keuning, and Johan Jeuring.
\newblock {N}ext-{S}tep {H}int {G}eneration for {I}ntroductory {P}rogramming
  {U}sing {L}arge {L}anguage {M}odels.
\newblock In {\em Proceedings of the Australasian Computing Education
  Conference (ACE)}, 2024.

\bibitem{DBLP:journals/corr/abs-2309-00029}
Natalie Kiesler, Dominic Lohr, and Hieke Keuning.
\newblock {E}xploring the {P}otential of {L}arge {L}anguage {M}odels to
  {G}enerate {F}ormative {P}rogramming {F}eedback.
\newblock {\em CoRR}, abs/2309.00029, 2023.

\bibitem{denny2024computing}
Paul Denny, James Prather, Brett~A. Becker, James Finnie-Ansley, Arto Hellas,
  Juho Leinonen, Andrew Luxton-Reilly, Brent~N. Reeves, Eddie~Antonio Santos,
  and Sami Sarsa.
\newblock {C}omputing {E}ducation in the {E}ra of {G}enerative {AI}.
\newblock {\em Communications of ACM}, 67(2), 2024.

\bibitem{DBLP:conf/icer/PhungPCGKMSS22}
Tung Phung, Victor{-}Alexandru Padurean, Jos{\'{e}} Cambronero, Sumit Gulwani,
  Tobias Kohn, Rupak Majumdar, Adish Singla, and Gustavo Soares.
\newblock {G}enerative {AI} for {P}rogramming {E}ducation: {B}enchmarking
  {C}hat{GPT}, {GPT}-4, and {H}uman {T}utors.
\newblock In {\em Proceedings of the Conference on International Computing
  Education Research - Volume 2 (ICER V.2)}, 2023.

\bibitem{edm23-pyfixv}
Tung Phung, Jos{\'{e}} Cambronero, Sumit Gulwani, Tobias Kohn, Rupak Majumdar,
  Adish Singla, and Gustavo Soares.
\newblock {G}enerating {H}igh-{P}recision {F}eedback for {P}rogramming {S}yntax
  {E}rrors using {L}arge {L}anguage {M}odels.
\newblock In {\em Proceedings of the International Conference on Educational
  Data Mining (EDM)}, 2023.

\bibitem{DBLP:conf/nips/PiechBHGSGS15}
Chris Piech, Jonathan Bassen, Jonathan Huang, Surya Ganguli, Mehran Sahami,
  Leonidas~J. Guibas, and Jascha Sohl{-}Dickstein.
\newblock {D}eep {K}nowledge {T}racing.
\newblock In {\em Proceedings of the Annual Conference on Neural Information
  Processing Systems (NeurIPS)}, 2015.

\bibitem{DBLP:conf/aaai/WuMGP19}
Mike Wu, Milan Mosse, Noah~D. Goodman, and Chris Piech.
\newblock {Z}ero {S}hot {L}earning for {C}ode {E}ducation: {R}ubric {S}ampling
  with {D}eep {L}earning {I}nference.
\newblock In {\em Proceedings of the {AAAI} Conference on Artificial
  Intelligence (AAAI)}, 2019.

\bibitem{DBLP:conf/nips/NieBP21}
Allen Nie, Emma Brunskill, and Chris Piech.
\newblock {P}lay to {G}rade: {T}esting {C}oding {G}ames as {C}lassifying
  {M}arkov {D}ecision {P}rocess.
\newblock In {\em Proceedings of the Annual Conference on Neural Information
  Processing Systems (NeurIPS)}, 2021.

\bibitem{DBLP:conf/edm/WuDDPG20}
Mike Wu, Richard~Lee Davis, Benjamin~W. Domingue, Chris Piech, and Noah~D.
  Goodman.
\newblock {V}ariational {I}tem {R}esponse {T}heory: {F}ast, {A}ccurate, and
  {E}xpressive.
\newblock In {\em Proceedings of the International Conference on Educational
  Data Mining (EDM)}, 2020.

\bibitem{DBLP:conf/lak/KimP23}
Yunsung Kim and Chris Piech.
\newblock {T}he {S}tudent {Z}ipf {T}heory: {I}nferring {L}atent {S}tructures in
  {O}pen-{E}nded {S}tudent {W}ork {T}o {H}elp {E}ducators.
\newblock In {\em Proceedings of the International Learning Analytics and
  Knowledge Conference (LAK)}, 2023.

\bibitem{DBLP:conf/lats/WangSLP17}
Lisa Wang, Angela Sy, Larry Liu, and Chris Piech.
\newblock {D}eep {K}nowledge {T}racing {O}n {P}rogramming {E}xercises.
\newblock In {\em Proceedings of the Conference on Learning @ Scale (L@S)},
  2017.

\bibitem{DBLP:conf/sigcse/PiechMJS21}
Christopher Piech, Ali Malik, Kylie Jue, and Mehran Sahami.
\newblock {C}ode in {P}lace: {O}nline {S}ection {L}eading for {S}calable
  {H}uman-{C}entered {L}earning.
\newblock In {\em Proceedings of the Technical Symposium on Computer Science
  Education (SIGCSE)}, 2021.

\bibitem{edm22-ai-teacher-test}
Ana{\"{\i}}s Tack and Chris Piech.
\newblock {T}he {AI} {T}eacher {T}est: {M}easuring the {P}edagogical {A}bility
  of {B}lender and {GPT-3} in {E}ducational {D}ialogues.
\newblock In {\em Proceedings of the International Conference on Educational
  Data Mining (EDM)}, 2022.

\bibitem{DBLP:conf/lats/MarkelOLP23}
Julia~M. Markel, Steven~G. Opferman, James~A. Landay, and Chris Piech.
\newblock {GPT}each: {I}nteractive {TA} {T}raining with {GPT}-based {S}tudents.
\newblock In {\em Proceedings of the Conference on Learning @ Scale (L@S)},
  2023.

\bibitem{DBLP:conf/chi/LuFH0W23}
Xinyi Lu, Simin Fan, Jessica Houghton, Lu~Wang, and Xu~Wang.
\newblock {ReadingQuizMaker}: {A} {H}uman-{NLP} {C}ollaborative {S}ystem that
  {S}upports {I}nstructors to {D}esign {H}igh-{Q}uality {R}eading {Q}uiz
  {Q}uestions.
\newblock In {\em Proceedings of the CHI Conference on Human Factors in
  Computing Systems (CHI)}, 2023.

\bibitem{DBLP:conf/naacl/WangFHW22}
Xu~Wang, Simin Fan, Jessica Houghton, and Lu~Wang.
\newblock {T}owards {P}rocess-{O}riented, {M}odular, and {V}ersatile {Q}uestion
  {G}eneration that {M}eets {E}ducational {N}eeds.
\newblock In {\em Proceedings of the Conference of the North American Chapter
  of the Association for Computational Linguistics (NAACL)}, 2022.

\bibitem{DBLP:conf/acl/Hua022}
Xinyu Hua and Lu~Wang.
\newblock {E}fficient {A}rgument {S}tructure {E}xtraction with {T}ransfer
  {L}earning and {A}ctive {L}earning.
\newblock In {\em Findings of the Association for Computational Linguistics
  (Findings of ACL)}, 2022.

\bibitem{DBLP:conf/emnlp/ZhangKLLL023}
Yunxiang Zhang, Muhammad Khalifa, Lajanugen Logeswaran, Moontae Lee, Honglak
  Lee, and Lu~Wang.
\newblock {M}erging {G}enerated and {R}etrieved {K}nowledge for {O}pen-{D}omain
  {QA}.
\newblock In {\em Proceedings of the Conference on Empirical Methods in Natural
  Language Processing (EMNLP)}, 2023.

\bibitem{DBLP:journals/corr/abs-2310-19208}
Xin Liu, Muhammad Khalifa, and Lu~Wang.
\newblock {L}it{C}ab: {L}ightweight {C}alibration of {L}anguage {M}odels on
  {O}utputs of {V}aried {L}engths.
\newblock {\em CoRR}, abs/2310.19208, 2023.

\bibitem{neurips2023gaied_3_guo}
Zhen Guo and Shangdi Yu.
\newblock {AuthentiGPT: Detecting Machine-Generated Text via Black-Box Language
  Models Denoising}.
\newblock {\em NeurIPS'23 Workshop on Generative AI for Education (GAIED)},
  2023.

\bibitem{neurips2023gaied_4_jahanbakhsh}
Kazem Jahanbakhsh, Mahdi Hajiabadi, Vipul Gagrani, Jennifer Louie, and Saurabh
  Khanwalkar.
\newblock {Beyond Hallucination: Building a Reliable Question Answering \&
  Explanation System with GPTs}.
\newblock {\em NeurIPS'23 Workshop on Generative AI for Education (GAIED)},
  2023.

\bibitem{neurips2023gaied_5_koutcheme}
Charles Koutcheme, Nicola Dainese, Sami Sarsa, Juho Leinonen, Arto Hellas, and
  Paul Denny.
\newblock {Benchmarking Educational Program Repair}.
\newblock {\em NeurIPS'23 Workshop on Generative AI for Education (GAIED)},
  2023.

\bibitem{neurips2023gaied_7_padurean}
Victor-Alexandru P{\u a}durean, Georgios Tzannetos, and Adish Singla.
\newblock {Neural Task Synthesis for Visual Programming}.
\newblock {\em NeurIPS'23 Workshop on Generative AI for Education (GAIED)},
  2023.

\bibitem{neurips2023gaied_8_lee}
Unggi Lee, Sanghyeok Lee, Junbo Koh, Yeil Jeong, Haewon Jung, Gyuri Byun,
  Yunseo Lee, Jewoong Moon, Jieun Lim, and Hyeoncheol Kim.
\newblock {Generative Agent for Teacher Training: Designing Educational
  Problem-Solving Simulations with Large Language Model-based Agents for
  Pre-Service Teachers}.
\newblock {\em NeurIPS'23 Workshop on Generative AI for Education (GAIED)},
  2023.

\bibitem{neurips2023gaied_9_blobstein}
Ariel Blobstein, Daniel Izmaylov, Tal Yifat, Michal Levy, and Avi Segal.
\newblock {Angel: A New Generation Tool for Learning Material based Questions
  and Answers}.
\newblock {\em NeurIPS'23 Workshop on Generative AI for Education (GAIED)},
  2023.

\bibitem{neurips2023gaied_10_mcnichols}
Hunter McNichols, Wanyong Feng, Jaewook Lee, Alexander Scarlatos, Digory Smith,
  Simon Woodhead, and Andrew Lan.
\newblock {Automated Distractor and Feedback Generation for Math
  Multiple-choice Questions via In-context Learning}.
\newblock {\em NeurIPS'23 Workshop on Generative AI for Education (GAIED)},
  2023.

\bibitem{neurips2023gaied_11_zhang}
Zihao Zhang, Zonghai Yao, Huixue Zhou, Feiyun Ouyang, and Hong Yu.
\newblock {EHRTutor: Enhancing Patient Understanding of Discharge
  Instructions}.
\newblock {\em NeurIPS'23 Workshop on Generative AI for Education (GAIED)},
  2023.

\bibitem{neurips2023gaied_12_hayat}
Ahatsham Hayat and Mohammad~Rashedul Hasan.
\newblock {Personalization and Contextualization of Large Language Models For
  Improving Early Forecasting of Student Performance}.
\newblock {\em NeurIPS'23 Workshop on Generative AI for Education (GAIED)},
  2023.

\bibitem{neurips2023gaied_14_sonkar}
Shashank Sonkar, MyCo Le, Xinghe Chen, Naiming Liu, Debshila~Basu Mallick, and
  Richard Baraniuk.
\newblock {Code Soliloquies for Accurate Calculations in Large Language
  Models}.
\newblock {\em NeurIPS'23 Workshop on Generative AI for Education (GAIED)},
  2023.

\bibitem{neurips2023gaied_15_savelka}
Jaromir Savelka, Paul Denny, Mark~H. Liffiton, and Brad~E Sheese.
\newblock {Efficient Classification of Student Help Requests in Programming
  Courses Using Large Language Models}.
\newblock {\em NeurIPS'23 Workshop on Generative AI for Education (GAIED)},
  2023.

\bibitem{neurips2023gaied_16_pielage}
Leon Pielage, Paul Schmidle, Bernhard Marschall, and Benjamin Risse.
\newblock {Diffusion Models in Dermatological Education: Flexible High Quality
  Image Generation for VR-based Clinical Simulations}.
\newblock {\em NeurIPS'23 Workshop on Generative AI for Education (GAIED)},
  2023.

\bibitem{neurips2023gaied_17_hwang}
Kevin~Timothy Hwang, Sai Challagundla, Maryam~M Alomair, Lujie~Karen Chen, and
  Fow-Sen Choa.
\newblock {Towards AI-Assisted Multiple Choice Question Generation and Quality
  Evaluation at Scale: Aligning with Bloom’s Taxonomy}.
\newblock {\em NeurIPS'23 Workshop on Generative AI for Education (GAIED)},
  2023.

\bibitem{neurips2023gaied_18_fawzi}
Fares Fawzi, Sadie Amini, and Sahan Bulathwela.
\newblock {Small Generative Language Models for Educational Question
  Generation}.
\newblock {\em NeurIPS'23 Workshop on Generative AI for Education (GAIED)},
  2023.

\bibitem{neurips2023gaied_19_han}
Jieun Han, Haneul Yoo, Junho Myung, Minsun Kim, Tak~Yeon Lee, So-Yeon Ahn, and
  Alice Oh.
\newblock {Exploring Student-ChatGPT Dialogue in EFL Writing Education}.
\newblock {\em NeurIPS'23 Workshop on Generative AI for Education (GAIED)},
  2023.

\bibitem{neurips2023gaied_21_anand}
Avinash Anand, Mohit Gupta, Kritarth Prasad, Navya Singla, Sanjana Sanjeev,
  Jatin Kumar, Adarsh~Raj Shivam, and Rajiv~Ratn Shah.
\newblock {Mathify: Evaluating Large Language Models on Mathematical Problem
  Solving Tasks}.
\newblock {\em NeurIPS'23 Workshop on Generative AI for Education (GAIED)},
  2023.

\bibitem{neurips2023gaied_26_oli}
Priti Oli, Rabin Banjade, Jeevan Chapagain, and Vasile Rus.
\newblock {The Behavior of Large Language Models When Prompted to Generate Code
  Explanations}.
\newblock {\em NeurIPS'23 Workshop on Generative AI for Education (GAIED)},
  2023.

\bibitem{neurips2023gaied_27_lira}
Hernan Lira, Luis Mart\'{i}, and Nayat~S\'{a}nchez Pi.
\newblock {Enhancing Writing Skills of Chilean Adolescents: Assisted Story
  Creation with LLMs}.
\newblock {\em NeurIPS'23 Workshop on Generative AI for Education (GAIED)},
  2023.

\bibitem{neurips2023gaied_30_ghadban}
Yasmina~Al Ghadban, Huiqi~(Yvonne) Lu, Uday Adavi, Ankita Sharma, Sridevi Gara,
  Neelanjana Das, Bhaskar Kumar, Renu Johns, Praveen Devarsetty, and Jane~E.
  Hirst.
\newblock {Transforming Healthcare Education: Harnessing Large Language Models
  for Frontline Health Worker Capacity Building using Retrieval-Augmented
  Generation}.
\newblock {\em NeurIPS'23 Workshop on Generative AI for Education (GAIED)},
  2023.

\bibitem{neurips2023gaied_31_asthana}
Sumit Asthana, Taimoor Arif, and Kevyn Collins-Thompson.
\newblock {Field Experiences and Reflections on Using LLMs to Generate
  Comprehensive Lecture Metadata}.
\newblock {\em NeurIPS'23 Workshop on Generative AI for Education (GAIED)},
  2023.

\bibitem{neurips2023gaied_32_zamfirescu-pereira}
J.D. Zamfirescu-Pereira, Laryn Qi, Bjorn Hartmann, John DeNero, and Narges
  Norouzi.
\newblock {Conversational Programming with LLM-Powered Interactive Support in
  an Introductory Computer Science Course}.
\newblock {\em NeurIPS'23 Workshop on Generative AI for Education (GAIED)},
  2023.

\bibitem{neurips2023gaied_33_bailis}
Suma Bailis, Lara McConnaughey, Jane Friedhoff, Feiyang Chen, Chase Adams, and
  Jacob Moon.
\newblock {WordPlay: An Agent Framework for Language Learning Games}.
\newblock {\em NeurIPS'23 Workshop on Generative AI for Education (GAIED)},
  2023.

\bibitem{neurips2023gaied_34_choi}
Jun~Ho Choi, Oliver Garrod, Paul Atherton, Andrew Joyce-Gibbons, Miriam
  Mason-Sesay, and Daniel Bj\"{o}rkegren.
\newblock {Are LLMs Useful in the Poorest Schools? TheTeacher.AI in Sierra
  Leone}.
\newblock {\em NeurIPS'23 Workshop on Generative AI for Education (GAIED)},
  2023.

\bibitem{neurips2023gaied_35_abdelghani}
Rania Abdelghani, H\'{e}l\`{o}ne Sauz\'{e}on, and Pierre-Yves Oudeyer.
\newblock {Generative AI in the Classroom: Can Student Remain Active Learners?}
\newblock {\em NeurIPS'23 Workshop on Generative AI for Education (GAIED)},
  2023.

\bibitem{neurips2023gaied_38_schmucker}
Robin Schmucker, Meng Xia, Amos Azaria, and Tom Mitchell.
\newblock {Ruffle\&Riley: Towards the Automated Induction of Conversational
  Tutoring Systems}.
\newblock {\em NeurIPS'23 Workshop on Generative AI for Education (GAIED)},
  2023.

\bibitem{neurips2023gaied_39_bulusu}
Arya Bulusu, Brandon Man, Ashish Jagmohan, Aditya Vempaty, and Jennifer
  Mari-Wyka.
\newblock {An Automated Graphing System for Mathematical Pedagogy}.
\newblock {\em NeurIPS'23 Workshop on Generative AI for Education (GAIED)},
  2023.

\bibitem{neurips2023gaied_40_levonian}
Zachary Levonian, Chenglu Li, Wangda Zhu, Anoushka Gade, Owen Henkel,
  Millie-Ellen Postle, and Wanli Xing.
\newblock {Retrieval-augmented Generation to Improve Math Question-Answering:
  Trade-offs Between Groundedness and Human Preference}.
\newblock {\em NeurIPS'23 Workshop on Generative AI for Education (GAIED)},
  2023.

\bibitem{neurips2023gaied_41_lekan}
Kasra Lekan and Zachary~A. Pardos.
\newblock {AI-Augmented Advising: A Comparative Study of ChatGPT-4 and
  Advisor-based Major Recommendations}.
\newblock {\em NeurIPS'23 Workshop on Generative AI for Education (GAIED)},
  2023.

\bibitem{neurips2023gaied_43_xu}
Austin Xu, Klinton Bicknell, and Will Monroe.
\newblock {Large Language Model Augmented Exercise Retrieval for Personalized
  Language Learning}.
\newblock {\em NeurIPS'23 Workshop on Generative AI for Education (GAIED)},
  2023.

\bibitem{neurips2023gaied_44_bhandari}
Shreya Bhandari, Yunting Liu, and Zachary~A. Pardos.
\newblock {Evaluating ChatGPT-generated Textbook Questions using IRT}.
\newblock {\em NeurIPS'23 Workshop on Generative AI for Education (GAIED)},
  2023.

\bibitem{neurips2023gaied_46_sahai}
Shubham Sahai, Umair~Z. Ahmed, and Ben Leong.
\newblock {Improving the Coverage of GPT for Automated Feedback on High School
  Programming Assignments}.
\newblock {\em NeurIPS'23 Workshop on Generative AI for Education (GAIED)},
  2023.

\bibitem{neurips2023gaied_47_roy}
Anirban Roy, Sujeong Kim, Claire Christensen, and Madeline Cincebeaux.
\newblock {Detecting Educational Content in Online Videos by Combining
  Multimodal Cues}.
\newblock {\em NeurIPS'23 Workshop on Generative AI for Education (GAIED)},
  2023.

\bibitem{neurips2023gaied_48_acun}
Cagla Acun and Ramazan Acun.
\newblock {GAI-Enhanced Assignment Framework: A Case Study on Generative AI
  Powered History Education}.
\newblock {\em NeurIPS'23 Workshop on Generative AI for Education (GAIED)},
  2023.

\end{thebibliography}
\bibliographystyle{unsrt}

\end{document}